\documentclass[a4wide,12pt]{article}

\usepackage{amssymb}
\usepackage{amsmath,amssymb,amsthm,amsfonts}
\usepackage{hyperref}
\usepackage{graphicx,color,colortbl}
\usepackage{upgreek}
\usepackage[mathscr]{euscript}
\usepackage{hhline}
\DeclareMathAlphabet{\mathpzc}{OT1}{pzc}{m}{it}
\usepackage{mathbbol,mathtools,calc}
\usepackage[labelfont={rm}]{subfig}
\usepackage{bm}
\usepackage{framed}

\usepackage{blkarray}

\DeclareMathAlphabet{\mathpzc}{OT1}{pzc}{m}{it}
\DeclareMathAlphabet{\mathdutchcal}{U}{dutchcal}{m}{n}
\SetMathAlphabet{\mathdutchcal}{bold}{U}{dutchcal}{b}{n}

\allowdisplaybreaks
\numberwithin{equation}{section}
\usepackage{ytableau}
\usepackage{mathdots}
\usepackage{tikz}
\usetikzlibrary{
	shapes,
	arrows,
	positioning,
	decorations.markings,
	decorations.pathmorphing,
	circuits.logic.US,
	circuits.logic.IEC,
	fit,
	calc,
	plotmarks,
	matrix
}

\tikzset{
>=stealth',
help lines/.style={dashed, thick},
axis/.style={<->},
important line/.style={thick},
connection/.style={thick, dotted},
punkt/.style={
rectangle,
rounded corners,
draw=black, thick,
text width=4.5em,
minimum height=2em,
text centered,
},
pil/.style={
->,
thick,
gray,
shorten <=2pt,
shorten >=2pt,}
}

\usepackage{array}
\usepackage{adjustbox}
\usepackage{cleveref}
\usepackage{enumitem}

\usepackage[tmargin=3cm,lmargin=3cm,rmargin=3cm,bmargin=3cm,headheight=1cm,footskip=1.5cm]{geometry}

\usepackage{fancyhdr}

\newtheorem*{theorem*}{Theorem}

\theoremstyle{definition}

\newtheorem*{remark*}{Remark}

\begin{document}
\title{Exact solutions to macroscopic fluctuation theory through classical integrable systems}

\author{
\vspace{5mm}
Kirone Mallick\footnote {~E-mail: kirone.mallick@ipht.fr},
Hiroki Moriya\footnote {~E-mail: hiroki.moriya@sorbonne-universite.fr}, 
Tomohiro Sasamoto\footnote {~E-mail: sasamoto@phys.titech.ac.jp}
\\
{\it $^*$Institut de Physique Th\'eorique, CEA, CNRS, Universit\'e Paris--Saclay,}\\
{\it  F--91191 Gif-sur-Yvette cedex, France, }\\
{\it $^{\dag}$Laboratoire de Physique Th\'eorique de la Mati\`ere Condens\'ee, }\\
{\it Sorbonne Universit\'e, 4 Place Jussieu, 75005 Paris, France,}\\
{\it $^{\ddag}$Department of Physics, Tokyo Institute of Technology,}\\
\vspace{5mm}
{\it Oh-okayama 2-12-1, Meguro-ku, Tokyo 152-8551, Japan}\\
}

%
%
%


%
%

\date{}

\maketitle

\begin{abstract}

  We give a short overview of recent developments in
  exact solutions for macroscopic fluctuation theory by using 
connections to classical integrable systems. A calculation of the cumulant generating function for a tagged particle 
is also given, agreeing with a previous result obtained from a  microscopic analysis. 

\end{abstract}


\section{Introduction}
\label{sec:Intro}

Fluctuations of non-equilibrium many-body systems have been a central subject in statistical physics for a long time.  Whereas 
equilibrium systems are described by  the  Gibbs measure allowing us to 
clarify  many intriguing and 
universal  thermodynamic properties, no  standard prescription for the stationary measure of 
non-equilibrium processes  is  available. Indeed, the  study of
stationary and time-dependent fluctuations
out of equilibrium remains a very challenging problem. 
 
There are several types of fluctuations on different time scales. This may be exemplified by a simple random walker
on a  one dimensional lattice, which starts at the origin  and  hops
 symmetrically to the left or  to the right neighboring 
sites with  probability  1/2. At time step $n$, the average and variance of the random walker are given by 
$0$ and $n$ respectively. The latter implies that the typical fluctuation of the position of the walker is of order  $O(n^{1/2})$. 
Then, according to the central limit theorem, this fluctuation
on the scale $O(n^{1/2})$  is given by the Gaussian distribution.


We can also study  the probability that the position
of the random walker is on  a  scale of $O(n)$. 
Since  typical fluctuations are of order  $O(n^{1/2})$, such a probability should vanish when $n$ tends to infinity. 
Nevertheless, one can  investigate how this probability decays to 0 when $n$ is large; 
it takes the form, $\mathbb{P}[X_n = n y] \simeq e^{-\Phi(y) n}$ where $X_n$ is the 
position of the walker at time $n$. 
This type of fluctuation is called a large deviation and the function $\Phi(y)$
is known as the rate function we wish to determine, in general. 
For the random walker the rate function is simply given by the binary entropy. 

Large deviation theory  is a huge subfield  of probability \cite{Varadhan1984,dembo2009large, Touchette2009}  and its  close connection to  equilibrium statistical 
mechanics is  known \cite{ellis2006entropy}. More recently, large deviation theory
has become one of the essential tools to study 
fluctuations of non-equilibrium systems.  
For instance,  fluctuation theorems, which describe a symmetry property of generic non-equilibrium systems
can be  formulated in the language of large deviation \cite{GallavottiCohen1995, LebowitzSpohn1999}. 

Studying large deviation properties of  a many particles system is a  difficult
problem.
Around 2000, Bertini, De Sole, Gabrielli, Jona-Lasinio and Landim
developed a   framework,  called {\it the  macroscopic fluctuation theory} (MFT) \cite{Bertini_2002},  to analyze  large deviations. 
Their  original scheme was formulated 
for diffusive systems and has been utilized to understand various properties of such processes \cite{Bertini_2015}. 
It should be remarked that  large deviations for certain stochastic models had been considered earlier:
for example, the large deviation principle for the symmetric simple exclusion process  (SEP) 
was  established in \cite{Ki1989}.  Besides, in  recent  years,
 various extensions
have been proposed, such as the  ballistic MFT \cite{Doyon_2023} and a quantum
version of  MFT \cite{Bernard_2021}.

The    macroscopic fluctuation theory is based on a variational principle
and the  associated Euler-Lagrange equations,  called hereafter
  {\it the  MFT equations}, play a key role. 
These equations govern the large deviation properties of the system  but 
they are often difficult to solve exactly because, in most cases,
they are coupled nonlinear partial differential equations (PDEs). 
Until recently, solutions to the MFT equations had been available only for some stationary 
cases and for the case of independent particles. 

In the last few years, there has been a significant 
progress for studying time dependent large deviation for interacting systems. 
This has been made possible thanks to the  discovery
of subtle connections between the  MFT equations and
some classical integrable systems. 
The purpose of this article is to give a  brief  account of these recent developments,
preceded by some explanations about the  basics of MFT. 
Our discussion of  classical integrable systems will be mostly based
on \cite{Mallick_2022}. We shall also extend our previous work by calculating, within
the MFT framework,  the  large deviation of a tagged particle in the
symmetric simple exclusion process; this purely macroscopic derivation will be shown to be
identical to the microscopic formula that was obtained earlier by Bethe Ansatz \cite{IMS2021}.

The rest of the paper is organized as follows. In section \ref{sec:models} we introduce a few models such as the SEP, 
reflective Brownian motions (RBM) and  the Kipnis-Marchioro-Presutti (KMP) model. In section \ref{sec:MFT}, we review the 
basics of MFT. In section \ref{sec:RBM}, we  discuss  results for RBM. In section \ref{sec:MFTint}, we describe  a few cases where the 
MFT equations are directly related to classical integrable systems,
namely the KMP model and the Kardar-Parisi-Zhang (KPZ)
equation with weak noise. However, the SEP requires a special discussion because  the
corresponding MFT equations are not manifestly integrable.
In section \ref{sec:SEP}, we present  a  general mapping from the MFT equations 
to the integrable AKNS system, which works for all interacting
particles  models  with 
constant diffusivity and a quadratic mobility,  including SEP ;  basic ideas of  the inverse scattering method are then  explained. 
Some details of the calculations of the cumulant generating function for a tagged particle are given in section \ref{sec:genX}. 
The article ends with concluding remarks in section \ref{sec:conc}.

\section{Models and quantities} 
\label{sec:models}

\subsection{SEP}

\unitlength .6mm
\begin{figure}
\begin{picture}(200,50)(-40,-5)
\multiput(0,10)(0.5,0){20}{\line(1,0){0.3}}
\multiput(0,30)(0.5,0){20}{\line(1,0){0.3}}
\multiput(150,10)(0.5,0){20}{\line(1,0){0.3}}
\multiput(150,30)(0.5,0){20}{\line(1,0){0.3}}
\put(10,10){\line(1,0){140}}
\put(10,30){\line(1,0){140}}
\put(10,10){\line(0,1){20}}
\put(30,10){\line(0,1){20}}
\put(50,10){\line(0,1){20}}
\put(70,10){\line(0,1){20}}
\put(90,10){\line(0,1){20}}
\put(110,10){\line(0,1){20}}
\put(130,10){\line(0,1){20}}
\put(150,10){\line(0,1){20}}
\put(0,19){$\cdots$}
\put(40,20){\circle*{10}}
\put(46,18){{\Large $\Rightarrow$}}
\put(48,33){$1$}
\put(24,18){{\Large $\Leftarrow$}}
\put(28,33){$1$}
\put(64,18){{\Large $\Leftarrow$}}
\put(68,33){$1$}
\put(80,20){\circle*{10}}
\put(100,20){\circle*{10}}
\put(106,18){{\Large $\Rightarrow$}}
\put(108,33){$1$}
\put(124,18){{\Large $\Leftarrow$}}
\put(128,33){$1$}
\put(140,20){\circle*{10}}
\put(153,19){$\cdots$}
\put(17,0){-3}
\put(37,0){-2}
\put(57,0){-1}
\put(79,0){0}
\put(99,0){1}
\put(119,0){2}
\put(139,0){3}
\end{picture}
\caption{SEP}
\label{fig:SEP}
\end{figure} 

In the symmetric simple exclusion process (SEP),  particles on a one-dimensional
lattice attempt  to hop from their current  position 
to  a neighboring site  with probability $dt$ during an
infinitesimal short duration $dt$. 
Because of hard-core  volume exclusion,
if the target site is already
occupied, the transition is forbidden (see Fig. \ref{fig:SEP}). 
A configuration of  particles of SEP at time $t$ is specified
by the binary variables $\eta_i(t)$ for all $i\in\mathbb{Z}$, where 
$\eta_i(t) =1 $  if the site $i$ is 
occupied and   $\eta_i(t) =0 $  otherwise.
The SEP has become one  of the most standard models
in non-equilibrium statistical physics 
and has been studied extensively \cite{Liggett1973, Spohn1991}. 
The unique translation invariant stationary measure of SEP is given by the Bernoulli measure in which all sites are 
independent and each site is occupied by a particle with probability $\rho$ where $0<\rho<1$. 
This parameter $\rho$ corresponds to the density of particles in SEP,  the only conserved quantity in  the model. 
The hydrodynamic behavior of the density profile is given by the diffusion equation \cite{KL1999}, as 
follows from the fact that the average density $\langle \eta_i(t)\rangle$ at site $i$ satisfies the lattice diffusion equation,
which can seen to be a consequence of the $su(2)$ symmetry of SEP \cite{Giardin__2009}. 
The large deviation principle for the density profile trajectory has been established in \cite{Ki1989}.  

In this article,  we shall focus on  the following quantities: the integrated current $Q_t$ at the bond (0,1) for time $[0,t]$,
which is defined as the total  number of particles which have hopped
from  site 0 to site 1 minus  the number of particles which
have jumped from   site 1 to site 0
during time interval $[0,t]$; we shall also consider
a related quantity, which  is the tagged particle position:  we assume
that at   initial time there is a particle at the origin, 
 we  put a tag on it and  we denote its position at time $t$ by $X_t$. 
The large deviation principle (LDP)
for these quantities was established in 2013 in \cite{SethuramanVaradhan2013}. 
Besides, for each site $x$,
we consider  the  `height' function $N(x,t)$,  defined as
\begin{equation}
N(x,t)=Q_t+
\begin{cases}
-\sum_{y=1}^x\eta_y(t), &\quad (x>0), \\
0,  &\quad (x=0), \\
+\sum_{y=x+1}^0\eta_y(t), &\quad (x<0). \\ 
\end{cases}
\end{equation}
Thanks to the
simple identity $\mathbb{P}[X_t  > x ] = \mathbb{P}[N(x,t) > 0]$, see
 the discussions in \cite{IMS2021}, 
 the study of  $X_t$ is equivalent to that of $N(x,t)$ (for all $x$).
 
For SEP, the rate function for the integrated current and the tagged particle position were calculated exactly by solving the model 
using Bethe Ansatz, combinatorial calculations and asymptotic analysis \cite{DG2009,IMS2021}.  One of the aims of the present work is to extract these results
 from MFT, at the macroscopic level.

\subsection{Reflective BM} 
One may consider a Brownian version of SEP, in which  many Brownian particles
reflect against  each other symmetrically 
when they collide. This  model, which can be obtained as
a low density limit of SEP, 
will be called 
the reflective Brownian motion (RBM).
 The uniform stationary measure is given by 
the Poisson point process in which distances of neighboring two particles are 
 independent Poisson distributions and   the 
 hydrodynamics is still governed by  the diffusion equation.
 The  integrated current $Q_t$, the position of the tagged particle $X_t$ and the height function  $N(x,t)$ can be defined and studied  in this model. 
Thanks to the symmetric nature of reflections, trajectories of the reflecting Brownian particles are statistically the same as 
those of  independent Brownian particles. Accordingly,
fluctuations of certain quantities
 in the  RBM,   such as the current,  can be related to those for 
independent Brownian particles.  Although  the  RBM model may  look elementary, the   
calculations of large deviations have 
some non-trivial and generic  aspects that
give very useful insights  for  studying interacting models
\cite{DG2009a, KMS2014, KMS2014, Sadhu_2015}. 



\subsection{KMP and related models} 
The  Kipnis-Marchioro-Presutti (KMP) model \cite{Kipnis:1982} is also  
defined on  a one-dimensional lattice, but now the
state at each site is a non-negative real variable $E_i$, which represents a quantity of  energy (or mass) present at  the site $i$. 
The stochastic time evolution rule is the following:  two neighboring sites $(i,i+1)$ are randomly selected and the sum $E_i+E_{i+1}$  of the energies is uniformly redistributed between  $i$ and $i+1$. 
The uniform stationary measure is given by independent exponential distribution $e^{-\beta E_i}$, which conforms to  the usual 
canonical distribution. 
The hydrodynamics is again given by the diffusion equation \cite{Kipnis:1982}, as  follows from the fact that the average density $\langle E_i(t) \rangle$ obeys the lattice diffusion equation (this 
can also  be understood as a consequence of stochastic duality due to the existence of $su(1,1)$ symmetry for the model
\cite{Giardin__2009}). 
In the last decade or so various models which have the same $su(1,1)$ symmetry and related to the KMP model have been introduced and studied
\cite{Frassek_2019a}.  
For the KMP and related models, one can study the integrated current $Q_t$ and the height function $N(x,t)$. 

SEP and  KMP  are cousin models: both  have a single conserved quantity,
the hydrodynamics  is the 
diffusion equation and some fluctuation properties can be studied in a similar manner. 
But there are also some important  differences. 
One technical point  is that for SEP the large deviation principle
has been established a long time ago whereas it has not yet  been proved for
KMP, although  some  exact but non-rigorous  results about large deviations
have been derived in \cite{Bertini_2005}. 
Another difference is that
for SEP many exact solutions using methods of quantum integrable
systems such as Bethe Ansatz are available. In contrast, 
the KMP model has not been ``exactly solved" at the microscopic level.
Nevertheless, some exact results 
for KMP, 
such as the hydrodynamics and stationary two point function,
have been found  by using the stochastic duality, 
that  follows from the $su(1,1)$ symmetry (but this does not imply integrability), see for instance \cite{CGRS2016}. 


\subsection{Typical fluctuations} 
In  the Introduction,
we recalled that  the typical fluctuations of a random walker are of order  $O(t^{1/2})$
with a Gaussian distribution. We shall now
 investigate the  typical fluctuations of the
current and  of the tagged particle position for the interacting 
 models discussed  above. 

For SEP, the typical fluctuation of a  tagged particle was studied a long time ago in \cite{A1983}. 
For the case of uniform density $\rho$, the average and the variance are given by 
\begin{align}
\langle X_t \rangle &= 0, \label{SEPtaga}\\
\langle X_T^2 \rangle &\simeq \frac{2(1-\rho)}{\rho}\sqrt{\frac{T}{\pi}}. \label{SEPtagv} 
\end{align}
Here,  and in the following,  $T$ means a final time which is considered to be large. 
The behavior of the variance indicates that the typical fluctuation of the tagged particle position of SEP is 
on the scale of $O(T^{1/4})$, which is much smaller than the case of a single random walker. 
It is further known that the distribution of the tagged particle position on this scale is Gaussian \cite{De_Masi_2002}. 
The results (\ref{SEPtaga}), (\ref{SEPtagv}) about typical fluctuations for the tagged particle position can be related  to 
those of the integrated current as 
\begin{align}
 \langle Q_t \rangle &= 0, \\
 \langle Q_T^2 \rangle &\simeq 2\rho(1-\rho)\sqrt{\frac{T}{\pi}} .  
\end{align}

Typical fluctuations for other models can also be studied. 
For the few models in the previous subsections, 
the average and the variance for the stationary measure with uniform density $\rho$ (particle density for SEP and RBM, 
or energy density for KMP) are summarized as \cite{Krapivsky_2012}
\begin{align}
 \langle Q_t \rangle &= 0, \\
 \langle Q_T^2 \rangle &\simeq \sigma(\rho) \sqrt{\frac{T}{\pi}} .
\end{align} 
Here $\sigma(\rho)$ is called the mobility and is given by 
\begin{equation}
 \sigma(\rho) =
 \begin{cases}
  2\rho(1-\rho), & \text{SEP,} \\
  2 \rho,            & \text{RBM,} \\
  \rho^2,            & \text{KMP}. 
  \end{cases}
\end{equation}
As explained in the next section, the MFT may be formulated for systems with given diffusivity $D(\rho)$ and mobility $\sigma(\rho)$.  
The diffusivity for all the three models above is a constant, equal to  1. 
The mobility which is quadratic in $\rho$ is often easier to handle. This case, including all the above three models,  may be treated 
in a unified way by introducing general quadratic $\sigma(\rho)=2A\rho(B-\rho)$ \cite{DG2009a}. 



\section{Macroscopic fluctuation theory}  
\label{sec:MFT}

\subsection{Coarse-grained description by Langevin equation}
We now describe  a coarse-grained description  of a microscopic model, with definite values of 
$D$ and $\sigma$,  such as SEP, RBM and KMP.
Since we are considering systems with
only one conserved quantity $\rho$, we 
can write a  continuity equation 
$\partial_t \rho + \partial_x j=0$. If the current $j$
obeys the Fourier's law or Fick's law, we have 
$j=-D \partial_x \rho$ with a constant $D$ and the continuity equation becomes the diffusion equation. This corresponds to the hydrodynamic
description of the models. To take into account the effects of fluctuation, we would add a noise to the current so that we put 
$j=-D \partial_x \rho + C \xi(x,t)$ where $\xi(x,t)$ is a Gaussian white noise with 
$\langle \xi(x,t)\rangle = 0, \langle \xi(x,t)\xi(x',t') \rangle =\delta(x-x')\delta(t-t')$. 
The strength of the noise $C$ should be taken to be consistent with the size of the fluctuation of the stationary measure
and is determined to be $C=\sqrt{\sigma(\rho)}$ \cite{Spohn1991}. 
We may also consider the case where the diffusion constant $D$ depends on the density $\rho$. 

To summarize, we may suppose that a coarse grained-description with 
fluctuations of a microscopic model is given by the Langevin equation, 
\begin{equation}
 \partial_t \rho = \partial_x [D(\rho) \partial_x \rho + \sqrt{\sigma(\rho)}\xi(x,t)] . 
\label{langevin} 
\end{equation}
Note that the height function at position $X$ and at time $T$ is given by 
\begin{equation}
N(X,T)=\int_{X}^\infty dx \rho(x,T) - \int_0^\infty dx \rho(x,0). 
\label{eq:height}
\end{equation}

\subsection{MFT action} 
Starting from the  Langevin equation (\ref{langevin}) and
using  a standard way to represent noise in terms of a functional integral  \cite{MSR1973},
the height generating function $\langle e^{\lambda N(X,T) } \rangle$
can be written as 
\begin{align}
\langle e^{\lambda N(X,T) } \rangle &= \int \mathcal{D}[\rho,H] e^{S[\rho,H]}, \label{fint} \\
\hbox{ with  } S[\rho,H] 
 &=  \lambda N(X,T)  -\mathcal{F}_0[\rho(x,0)]
  - \int_{0}^{T} dt \int_{-\infty}^\infty dx (H\partial_t \rho+
  \mathcal{H}),
\label{action}
\end{align}
where the  function $S$  is often called the MFT action. 
Here $N(X,T)$ is given  in terms of $\rho(x,t)$
using (\ref{eq:height})  and 
the  Hamiltonian $\mathcal{H}$ has the following expression
\begin{equation}
\mathcal{H}[\rho,H]
  =D(\rho)
  (\partial_x\rho)(\partial_xH)-\frac12\sigma(\rho)(\partial_xH)^2 \,.
  \label{mathcalHamil}
  \end{equation}
Finally,  the free energy of the initial density profile, chosen as a 
 local equilibrium state with density $\bar{\rho}(x)$, 
is given by
\begin{align}
\mathcal{F}_0[\rho(x,0)] &=  \int_{-\infty}^\infty dx
  \int_{\bar\rho(x)}^{\rho(x,0)} dr  \frac{2D(r)(\rho(x,0) - r)}{\sigma(r)}, \label{F0}
\end{align}

The derivation of these formulas  is  not very difficult. To obtain
 (\ref{fint}) one  starts from the identity 
\begin{align}
\label{gen1}
\langle e^{\lambda N(X,T) } \rangle 
&=
\langle  \int \mathcal{D}[\rho] e^{\lambda N(X,T)}  \delta(\partial_t \rho - \partial_x [D(\rho)\partial_x \rho + \sqrt{\sigma(\rho)}\xi(x,t)]) \rangle. 
\end{align}
Here $\delta$ is the Dirac delta function and the above expression simply means that we are summing over all possible 
density profile trajectories with the condition that the Langevin equation (\ref{langevin}) should be satisfied. 
Next we write the delta function as an integral and then, 
recalling that the noise $\xi$ is Gaussian with $\langle \xi(x,t)\rangle = 0, \langle \xi(x,t)\xi(x',t') \rangle =\delta(x-x')\delta(t-t')$,
we can take the average with respect to the noise, and obtain
(\ref{fint}). 
The term (\ref{F0}) comes from the initial condition, which is taken to be
 the local stationary measure with density $\bar{\rho}(x)$. 
The rate function for the density $\bar{\rho}$ is written in the form $f(\rho)-f(\bar{\rho})-(\rho-\bar{\rho})f'(\bar{\rho})$ where $f$ is the 
free energy density for the system with density $\rho$;
then using the fluctuation-dissipation theorem $f''(\rho)=2D(\rho)/\sigma(\rho)$, one finds (\ref{F0}) \cite{Derrida2007}.


\subsection{MFT equations} 
In principle the functional integral (\ref{fint}) contains the whole information of the system described by the 
Langevin equation (\ref{langevin}). 
Large $T$ behavior is dominated by the maximum of the action $S$ and is expected to be the same for the original microscopic models. 

Euler-Lagrange equations for the above MFT action are 
\begin{align}
\partial_t\rho &= \partial_x[D(\rho)\partial_x\rho- \sigma(\rho)\partial_xH], \label{MFTeq1}\\
\vspace{-3mm}
\partial_tH &= -D(\rho)\partial_x^2H- \sigma'(\rho)(\partial_xH)^2/2,  \label{MFTeq2} 
\end{align}
accompanied by the conditions at the initial and the final times:
\begin{align}
H(x,T) &= \lambda\theta(x-X), \label{MFTf}\\
H(x,0) &= \lambda\theta(x)+f'(\rho(x,0))-f'(\bar\rho(x)). \label{MFTi}
\end{align}
Here $f'(\rho) = \log \frac{\rho}{1-\rho}$ for SEP, $f'(\rho) =\log \rho$ for RBM and 
$f'(\rho) =-1/\rho$ for KMP model. 

These  coupled nonlinear partial differential equations(PDEs) are  in general difficult to solve. 
Most of the works until recently have been restricted to studies for stationary situations and for 
non-interacting models like RBM. Besides, perturbative expansions have also been performed \cite{Krapivsky_2012}.

\subsection{Stationary case} 
The first successes of the MFT were about the stationary case for systems with open boundaries. 
For SEP with open boundaries, the exact stationary measure can be constructed by the matrix product method \cite{DEHP1993}. 
Based on this representation the large deviation of the density profile was calculated in \cite{DLS2002}. 
At the same time,  the same large deviation was calculated
at the macroscopic level 
based on the MFT formulation \cite{Bertini_2003} (see also \cite{Tailleur_2008}). These results turned out to be the same: the microscopic and the  macroscopic results matched and this was  an important touchstone in the development of the MFT framework.
For other related works, see for instance \cite{Bodineau_2004, Derrida_2004}. 

\subsection{Reflective Brownian motion}
\label{sec:RBM}
The large deviation for time dependent case is more challenging. 
For RBM, however, even the time dependent large deviations are accessible in a rather simple manner  \cite{KMS2015}. 
Let us put the values of $D$ and $\sigma$ for RBM, i.e., $D=1, \sigma(\rho)=2\rho$ in (\ref{MFTeq1}) and (\ref{MFTeq2}). 
The MFT equations are still coupled non-linear equations, but applying the canonical Cole-Hopf transformation,
\begin{equation}
 Q=\rho e^{-H}, \quad P=e^H, 
\end{equation}
the equations are decoupled into  the diffusion and anti-diffusion equations, 
\begin{equation}
\partial_t Q = \partial_{xx} Q, ~ \partial_t P = -\partial_{xx} P.
\end{equation} 
These   linear equations are readily  solved  by Fourier transformations,
leading to explicit formulas for large deviation properties of  the  RBM \cite{KMS2015}.


\section{MFT equations and classical integrable systems}
\label{sec:MFTint}
\subsection{KMP} 
For the KMP model, the mobility is $\sigma(\rho)=\rho^2$ and the MFT equations read 
\begin{align}
\partial_t\rho &= \partial_x^2 \rho- \partial_x (\rho^2\partial_xH), \\
\partial_tH &= -\partial_x^2H-\rho(\partial_xH)^2. 
\end{align}
Equations for $\rho,\partial_x H$ are nothing but the derivative nonlinear Schr\"odinger equation \cite{Kaup1978}, 
which is a classical integrable system \cite{AS1981}. 
Exploiting this fact,  the MFT equations for the KMP model, with the initial condition $\rho(x)= \delta(x)$,  was solved in \cite{Bettelheim_2022}.
This may be considered as the first exact solution to the MFT equations in the time dependent regime for an interacting model. 
But,  because the KMP model has not been exactly solved at the microscopic level, 
 the validity of the obtained result can not be checked easily.

\subsection{Weak noise KPZ}
In this article,  we mainly focus on the MFT for diffusive systems. 
But similar equations also appear in related  systems, for example when one 
considers the  large deviation properties of the KPZ equation in the small noise limit.
It had been noticed in \cite{Janas_2016} that the optimal path equations
for the weak noise KPZ are in  fact integrable 
and a full  solution for  this case, using the Inverse Scattering Method,
was obtained in \cite{Krajenbrink_2021}.

\section{Solution to the MFT equation for SEP} 
\label{sec:SEP}
\subsection{Mapping to a classical integrable system}
\label{sec:transf}
For the case of SEP, the MFT equation itself is not a classical integrable system. 
In \cite{Mallick_2022}, we found a transformation of
the MFT equations for SEP to
a classical integrable system.  
The mapping  reads 
\begin{align}
u(x,t) &= \frac{1}{\sigma'(\rho)}\frac{\partial}{\partial x}\sigma(\rho)\exp\left[-\int_{-\infty}^x dy \sigma'(\rho)\partial_y H/2\right], \label{map1}
\\
v(x,t) &= -\frac{2}{\sigma'(\rho)}\frac{\partial}{\partial x}\exp\left[\int_{-\infty}^x dy\sigma'(\rho)\partial_yH/2\right]. \label{map2}
\end{align}
In fact the transformation works for any quadratic $\sigma$, and the MFT equations become 
\begin{align}
\partial_tu(x,t) &= \partial_{xx}u(x,t)-2u(x,t)^2v(x,t),  \\
\partial_tv(x,t) &= -\partial_{xx}v(x,t)+2u(x,t)v(x,t)^2.  
\label{AKNSeq}
\end{align}
They are still coupled non-linear PDEs but now they are exactly in the form of a well-known classical integrable system, 
known as the AKNS (Ablowitz-Kaup-Newell-Segur) system \cite{AS1981}. This opens up the possibility to study MFT equations 
of models with quadratic $\sigma$ using the standard machinery of classical integrable systems. 
Note that in the low density limit, in which $\sigma$ becomes simply $2\rho$, the $y$ integrals in the exponent of the transformation 
can be performed and they become equivalent to the canonical Cole-Hopf transformation. 

The above coupled PDEs should be solved under certain initial and final time conditions. 
In this article, we focus on the two sided stationary initial condition where $\bar{\rho}(x)=\rho_-\theta(-x)+\rho_+\theta(x)$. 
After the transformation, the conditions (\ref{MFTi}),(\ref{MFTf}) for the MFT equations become 
\begin{equation}
 u(x,0) = g_u \delta(x), ~ v(x,T) = g_v \delta(x-X)
 \label{uv_if}
\end{equation} 
with some coefficients $g_u$ and $g_v$. 
The function  $v$ becomes a $\delta$ function at final time
is related to the fact that we are interested in the integrated current. 
On the other hand,  the  condition that $u$ becomes proportional 
to a $\delta$ function at $t=0$ is
nontrivial and is very important in order  to solve the 
problem exactly. 
However, the amplitudes $g_u,g_v$ can not be determined from the transformation alone and must be fixed later. 
It had been known that the MFT equations for SEP may be mapped to the  Landau-Lifshitz equations (LLE)  for classical spin chains
\cite{Tailleur_2008},  
which are  classically integrable. However,
the initial and final time conditions 
for the LLE do not seem  easy to handle. 

The transformations (\ref{map1}),(\ref{map2}) were discovered while trying to find a  mapping from MFT equations of SEP to a tractable classical integrable system ; they  may  also be obtained by  combining some  known mappings between soliton equations \cite{Krajenbrink_2022}.

\subsection{Inverse scattering applied  to MFT} 
\label{ISideas}
Once the problem is mapped to a classical integrable system, one can try to use all available theoretical tools to derive exact solutions.
 Here, the  inverse scattering 
method (ISM) \cite{AS1981} turns out to be  particularly useful. 
The basic strategy of the ISM is to reformulate the non-linear PDE as
two auxiliary linear problems, one in space and the other in time. The first linear equation is 
regarded as a scattering problem, for which one can determine
the scattering amplitudes. The second linear equation 
ensures that the time evolution of these scattering amplitudes is simple
and fixes their values at any arbitrary time. 
Finally,  inverse scattering allows us to reconstruct  the original variables at the final time.

However, in our problem, there is  an  important difference from the usual
setup of the ISM.  Normally,  given the initial condition, we
 try to determine the variables at
an arbitrary time.  But, here, in the MFT case, 
 we do not have full information about the initial conditions. 
Rather,  we have one condition at the initial time and another at the final time. 
But, still, we can apply the basic strategy  of ISM to our problem. 
We may first solve the scattering problems and determine the scattering amplitudes 
both at initial and final times in terms of unknown functions. Since they have 
to be related by a simple time evolution, the unknown functions should satisfy 
some consistency conditions. In our case, these conditions can be written 
 as a scalar Riemann-Hilbert problem, which may be solved rather easily, leading to explicit
expressions of the  unknown functions, which, in turn, yield the 
 large deviation properties we are looking for. 

In the next section,  we apply these ideas to the calculation of the cumulant generating 
function of the height $N(X,t)$  for an arbitrary  value of the
position $X$. This is an extension  of  \cite{Mallick_2022}, in which only
the $X=0$ case was studied.

\section{Cumulant generating function for general $X$} 
\label{sec:genX}
\subsection{Linear problems} 
By following the standard procedures of inverse scattering method \cite{AS1981}, we  first consider the auxiliary linear problems, 
\begin{align}
\partial_x\Psi(x,t;k) &= U(x,t;k)\Psi(x,t;k), \label{eq:LPx} \\
\partial_t\Psi(x,t;k) &= V(x,t;k)\Psi(x,t;k), \label{eq:LPt}
\end{align}
where
\begin{align}
U(x,t;k) &= -ik\sigma_3+v\sigma_++u\sigma_-, \\
V(x,t;k) &= 2k^2\sigma_3+2ik[v\sigma_++u\sigma_-]+[uv\sigma_3-\partial_xv\sigma_++\partial_xu\sigma_-],
\end{align}
with $\sigma_3=(\begin{smallmatrix} 1&0\\0&-1\end{smallmatrix}), \sigma_+=(\begin{smallmatrix} 0&1\\0& 0\end{smallmatrix}),
\sigma_-=(\begin{smallmatrix} 0&0\\1&0\end{smallmatrix})$. 
The zero-curvature condition, $U_t-V_x+[U,V]=0$, which is the compatibility condition of the above two linear problems, 
gives the AKNS equation (\ref{AKNSeq}). This means that one may solve the original AKNS equation (\ref{AKNSeq}) by 
studying the two linear problems (\ref{eq:LPx}),(\ref{eq:LPt}). 

Because  $U(x,t)$ is vanishing as $x\to\pm\infty$,
 equation (\ref{eq:LPx}) may be thought of as a scattering problem. 
We may consider two solutions with the boundary conditions, 
\begin{alignat}{2}
&\phi(x;k)\sim
\begin{pmatrix}
1 \\
0
\end{pmatrix}
e^{-ikx}, \quad &\bar\phi(x;k)\sim
\begin{pmatrix}
0 \\
-1
\end{pmatrix}
e^{ikx}, \quad 
&\mathrm{as}\quad x\rightarrow-\infty, \label{eq:as-} \\
&\psi(x;k)\sim
\begin{pmatrix}
0 \\
1
\end{pmatrix}
e^{ikx}, \quad &\bar\psi(x;k)\sim
\begin{pmatrix}
1 \\
0
\end{pmatrix}
e^{-ikx}, \quad 
&\mathrm{as}\quad x\rightarrow+\infty, \label{eq:as+}
\end{alignat}
and define the scattering amplitudes $a(k),\bar a(k),b(k), \bar b(k)$ as 
coefficients of linear relations between them, 
\begin{align}
\phi(x;k) &= a(k)\bar\psi(x;k)+b(k)\psi(x;k), \label{eq:def_ab} \\
\bar\phi(x;k) &= \bar b(k)\bar\psi(x;k)-\bar a(k)\psi(x;k) \label{eq:def_barab}.
\end{align}
They satisfy the relation
\begin{equation}
 a(k) \bar{a}(k) + b(k) \bar{b}(k) = 1. 
\label{aabb1}
\end{equation}

\subsection{Scattering amplitudes in terms of  unknown functions} 
In our case, because the initial condition for $u$ is a $\delta$ function, one can solve the scattering problem easily,
see \cite{Mallick_2022}, as 
\begin{align}
a(k,0) &= 1+g_u\hat{v}_+(k), \quad b(k,0)=g_u, \label{eq:ISA1_ab} \\
\bar a(k,0) &= 1+g_u\hat{v}_-(k), \quad \bar b(k,0)=-\left[\hat{v}(k)+g_u\hat{v}_+(k)\hat{v}_-(k)\right],  \label{eq:ISA1_barab}
\end{align}
in terms of the unknown functions, 
\begin{equation}
\hat{v}_\pm(k) =\int_{\mathbb{R}_\pm}~v(x,0)e^{2ikx}dx
\end{equation}
and $\hat{v}(k):=\hat{v}_+(k)+\hat{v}_-(k)$, which are the half and full Fourier transforms of $v$ respectively. 

In the same way, because the final condition for $v$ is a $\delta$ function, one can solve the scattering problem easily as 
\begin{align}
a(k,T) &= 1+g_v\hat{u}_+(k), \quad b(k,T)=\left[\hat{u}(k)+g_v\hat{u}_+(k)\hat{u}_-(k)\right]e^{-2ikX}, \label{eq:FSA_ab} \\
\bar a(k,T) &= 1+g_v\hat{u}_-(k), \quad \bar b(k,T)=-g_ve^{2ikX} \label{eq:FSA_barab}, 
\end{align}
in terms of the unknown functions, 
\begin{equation}
\hat{u}_\pm(k):=\int_{\mathbb{R}_\mp}~u(x+X,T)e^{-2ikx}dx
\end{equation}
and $\hat{u}(k):=\hat{u}_+(k)+\hat{u}_-(k)$, which are the half and full Fourier transforms  of $u$ respectively. 
The functions $\hat{u}_\pm(k)$ are analytic in upper and lower complex half-plane.

In addition, thanks to $V(x,t;k)\sim2k^2\sigma_3$ for $x\to\pm\infty$, the linear problem for the time evolution (\ref{eq:LPt})  
is simplified and the time evolution of the scattering amplitudes become trivial. We have 
\begin{align}
a(k,t) &= a(k,0), \quad b(k,t)=b(k,0)e^{-4k^2t}, \label{eq:TE_ab} \\
\bar a(k,t) &= \bar a(k,0), \quad \bar b(k,t)=\bar b(k,0)e^{4k^2t}. \label{eq:TE_barab}
\end{align}

\subsection{Determination of the product $g_u g_v$}
Up to now, the weights of delta functions in the initial and final conditions (\ref{uv_if}) have been treated as general parameters. 
To express them in terms of original model parameters, let us consider the linear problem, 
\begin{equation}
\partial_x\Omega(x,t)=U(x,t;0)\Omega(x,t). 
\label{Omegak0}
\end{equation}
Noticing that this is nothing but the $k=0$ case of (\ref{eq:LPx}), the scattering amplitudes for $k=0$ can be written in terms of $\Omega$ as 
\begin{align}
 \lim_{\substack{x\to+\infty\\y\to-\infty}}\Omega(x,t)\Omega^{-1}(y,t) 
 = 
\begin{pmatrix}
a(0,t) & -\bar{b}(0,t) \\
b(0,t) & \bar{a}(0,t)
\end{pmatrix}.
\label{eq:TMab}
\end{align}
By setting $k=0$ in (\ref{eq:ISA1_ab}) and
(\ref{eq:FSA_barab}), we find $g_u=b(0,0)$ and $g_v=-\bar b(0,T)$. 
Besides, (\ref{Omegak0}) admits an explicit representation in terms of $(\rho,H)$ as 
\begin{equation}
\Omega(x,t)=
\begin{pmatrix}
e^{\int_{-\infty}^xdy~(1-\rho)\partial_yH} & e^{\int_{-\infty}^xdy~\rho\partial_yH} \\
-(1-\rho)e^{\int_{-\infty}^xdy~\rho\partial_yH} & \rho e^{\int_{-\infty}^xdy~(1-\rho)\partial_yH}
\end{pmatrix},
\end{equation}
as confirmed by direct checking. 
Taking into account the boundary conditions, $\rho\sim\rho_\pm$ and $H-\lambda/2\sim\pm\lambda/2$ for $x\to\pm\infty$, 
and using (\ref{eq:TMab}), we obtain
\begin{alignat}{2}
a(0,t) &= [1+(e^\lambda-1)\rho_-]e^{\Lambda/2-\lambda/2}, &\quad b(0,t) &= -(r_--e^{-\lambda}r_+)e^{-\Lambda/2+\lambda/2}, \label{eq:SV_ab} \\
\bar a(0,t) &= [1+(e^{-\lambda}-1)\rho_+]e^{-\Lambda/2+\lambda/2}, &\quad \bar b(0,t) &=(e^\lambda-1)e^{\Lambda/2-\lambda/2}, \label{eq:SV_barab}
\end{alignat}
with $\Lambda=\int_{-\infty}^{\infty} dx~(1-2\rho)\partial_xH$ being a conserved quantity and $r_\pm=\rho_\pm(1-\rho_\mp)$.
From this we conclude  that the product $g_u g_v$ is given, 
in terms of the  parameters  of the model,  by
\begin{equation}
g_ug_v=-b(0,t)\bar b(0,t)=(e^\lambda-1)\rho_-(1-\rho_+)+(e^{-\lambda}-1)\rho_+(1-\rho_-). 
\label{omega}
\end{equation}
The RHS of (\ref{omega}) is an important parameter and is denoted by $\omega$ in the following.

\subsection{Riemann-Hilbert problem} 
As mentioned in subsection \ref{ISideas}, the solution $u(x,T)$ at final time can be determined by using  the  consistency condition \eqref{eq:TE_ab} 
for  the scattering amplitudes,
given in \eqref{eq:ISA1_ab} and \eqref{eq:FSA_ab}. 
Writing down the relation (\ref{aabb1}) at time $T$ by using $b(k,T) = b(k,0)e^{-4k^2T} = g_u e^{-4k^2T}$, we find 
\begin{equation}
[1+g_v\hat{u}_+(k)][1+g_v\hat{u}_-(k)]=1 + g_u g_v e^{-4k^2T+2ikX}. 
\label{RHP}
\end{equation}
This is a scalar Riemann-Hilbert problem, where an analytic function is factorized into a product of functions analytic in the upper and lower half planes, 
and can be easily solved as 
\begin{equation}
1+g_v\hat u_\pm(k)=e^{Z_\pm(k)}
\end{equation}
where the  function $Z_\pm(k)$ is given by 
\begin{equation}
Z_\pm(k)=-\frac12\sum_{n=1}^\infty\frac{\left(-\omega e^{-4\left(k-\frac{iX}{4T}\right)^2T-X^2/4T}\right)^n}{n}\mathrm{erfc}\left[\mp i\sqrt{4nT}\left(k-\frac{iX}{4T}\right)\right].
\end{equation}
Note that a special value at $k=0$ is evaluated as 
\begin{equation}
\frac{Z_\pm'(0)}{(\pm2i)\sqrt{T}}=-\sum_{n=1}^\infty(-\omega)^n\left[\frac{e^{-n\xi^2}}{\sqrt{n\pi}}\pm\xi\mathrm{erfc}\left(\mp\sqrt{n}\xi\right)\right]
\label{eq:SV_Z}
\end{equation}
where $\xi = X/\sqrt{4T}$. 

The reduction of the problem to a scalar Riemann-Hilbert factorization is one of the main technical simplifications.  
A condition equivalent to (\ref{RHP}) was already obtained in  \cite{grabsch2022exact}, by an approach
 that did not  use inverse scattering.

\subsection{Cumulant generating function} 
The cumulant generating function is related to the height function by \cite{Bettelheim_2022b}
\begin{equation}
\sqrt{T}\mu'(\lambda)=N(X,T),
\label{eq:LT}
\end{equation}
which may be rewritten as 
\begin{equation}
\sqrt{T}\mu'(\lambda)=\int_X^\infty dx\left[\rho(x,T)-\rho_+\right]-\int_0^\infty dx\left[\rho(x,0)-\rho_+\right]-\rho_+X.
\label{eq:CGF_MFT}
\end{equation}
In terms of the ISM variables, this is written as 
\begin{equation}
\sqrt{T}\mu'(\lambda)=-\int_X^\infty dx\int_x^\infty dy~e^\Lambda u(y,T)-\rho_+(1-\rho_+)\int_0^\infty dx\int_x^\infty dy~e^{-\Lambda}v(y,0)-\rho_+X.
\label{eq:CGF_ISM}
\end{equation}
Rewriting in terms of the half Fourier transforms, we obtain
\begin{align}
\sqrt{T}\mu'(\lambda) &= -e^\Lambda\int_0^\infty dx~xu(x+X,T)-\rho_+(1-\rho_+)e^{-\Lambda}\int_0^\infty dx~xv(x,0)-\rho_+X \nonumber \\
 &= -e^\Lambda\frac{\hat{u}_-'(0)}{(-2i)}-\rho_+(1-\rho_+)e^{-\Lambda}\frac{\hat{v}_+'(0)}{(2i)}-\rho_+X \nonumber \\
 &= -e^\Lambda\frac{\bar a'(0)}{(-2i)g_v}-\rho_+(1-\rho_+)e^{-\Lambda}\frac{a'(0)}{(2i)g_u}-\rho_+X \nonumber \\
 &= -e^\Lambda\frac{Z_-'(0)\bar a(0)}{(-2i)g_v}-\rho_+(1-\rho_+)e^{-\Lambda}\frac{Z_+'(0)a(0)}{(2i)g_u}-\rho_+X.
\end{align}
In the third equality, we used 
the conserved quantities $a(k)=1+g_v\hat u_+(k)=1+g_u\hat v_+(k)=e^{Z_+(k)}$ and $\bar a(k)=1+g_v\hat u_-(k)=1+g_u\hat v_-(k)=e^{Z_-(k)}$.
This  shows that  the  cumulant generating function depends  only 
on  conserved quantities.
Plugging the special values \eqref{eq:SV_Z}, we obtain 
\begin{align}
\mu'(\lambda) &= -g_ue^\Lambda\bar a(0)\sum_{n=1}^\infty(-\omega)^{n-1}\left[\frac{e^{-n\xi^2}}{\sqrt{n\pi}}+\xi\mathrm{erf}\left(\sqrt{n}\xi\right)-\xi\right] \nonumber \\
 &\quad -\rho_+(1-\rho_+)g_ve^{-\Lambda}a(0)\sum_{n=1}^\infty(-\omega)^{n-1}\left[\frac{e^{-n\xi^2}}{\sqrt{n\pi}}+\xi\mathrm{erf}\left(\sqrt{n}\xi\right)+\xi\right]-2\rho_+\xi . \nonumber \\
\end{align}
Using \eqref{eq:SV_ab} and \eqref{eq:SV_barab}, we find 
\begin{equation}
-g_ue^\Lambda\bar a(0)-\rho_+(1-\rho_+)g_ve^{-\Lambda}a(0)= \omega'(\lambda)
\end{equation}
and
\begin{equation}
\frac{g_ue^\Lambda\bar a(0)-\rho_+(1-\rho_+)g_ve^{-\Lambda}a(0)}{1+\omega}-2\rho_+=-\frac{e^\lambda\rho_-}{1+(e^\lambda-1)\rho_-}-\frac{e^{-\lambda}\rho_+}{1+(e^{-\lambda}-1)\rho_+}.
\end{equation}
Therefore, integrating $\mu'(\lambda)$ w.r.t. $\lambda$, we conclude that the cumulant generating function is of the form:
\begin{equation}
\mu(\lambda)=-\sum_{n=1}^\infty\frac{(-\omega)^n}{n^{3/2}}\left[\frac{e^{-n\xi^2}}{\sqrt{\pi}}+\sqrt{n}\xi\mathrm{erf}\left(\sqrt{n}\xi\right)\right]-\xi\log\frac{1+(e^\lambda-1)\rho_-}{1+(e^{-\lambda}-1)\rho_+}.
\end{equation}
This agrees with the formula in \cite{IMS2021} which was found by directly studying the SEP microscopically.\footnote{
Possible appearance of solitons for general $x$ are not discussed in this paper because they are known not to change the 
final expression of the cumulant generating function for a few similar models \cite{Bettelheim_2022b}.} 

In this paper we focused on the calculation of the cumulant generating function, we can also study optimal profiles of $\rho$ and $H$ 
at initial and final times as in \cite{Mallick_2022}.

\section{Concluding Remarks} 
\label{sec:conc}
In this article, after recalling  some basic
facts about MFT, we  present  a recent exact solution 
of  the MFT equations, found  
by connecting them to a classical integrable system. We have also
performed an original  calculation of the cumulant generating function for 
the tagged particle position, which matches perfectly with
the formula obtained by microscopic calculations in \cite{IMS2021}. 

We also emphasize the fact that the transformation of \cite{Mallick_2022} can be applied to MFT equations for all models 
with quadratic mobility $\sigma(\rho)$ which include not only SEP but also KMP and related models. For SEP, the  
large deviations of the same physical observables (current, height, tagged particle position) 
have been derived  directly from microscopic models. The results agree 
with  the predictions  of MFT obtained by solving
the MFT equations. On the other hand for the KMP model, an  exact analysis of the microscopic model has not  yet 
been performed. Thus, the solution of
the MFT equation for the KMP model provides  an indispensable tool
to study large deviation of the 
model  and might also give further motivation to study the microscopic model directly and exactly. 

After the first few works on the subject, there have been several works in various directions 
\cite{Bettelheim_2022b,Schorlepp_2023,smith2023macroscopic,hartmann2023probing,krajenbrink2023weak,Krajenbrink_2023,tsai2022integrability,bettelheim2023whitham, bettelheim2024}. 
It now seems obvious that this new approach to study large deviations of interacting models will provide a  wider and deeper understanding 
of fluctuations of non-equilibrium systems.

\bigskip
\noindent
{\bf Acknowledgments.}
The authors thank Aur\'elien Grabsch, Pierre Rizkallah, Pierre Illien and Olivier B\'enichou for useful discussions on related problems. 
The work of KM has been supported by the project RETENU ANR-20-CE40-0005-01 of the French National Research Agency (ANR).
HM would like to acknowledge all the support of LPTMC and Sorbonne University.
Part of this article was written during stays of TS at the Isaac Newton Institute of Mathematical Sciences and at Institut des Hautes \'Etudes Scientifiques. 
The work of TS has been supported by JSPS KAKENHI Grants No. JP21H04432, No. JP22H01143.





\newcommand{\etalchar}[1]{$^{#1}$}

\end{document}